# Electronic-correlation-assisted charge stripe order in a Kagome superconductor

Linwei Huai[1,#], Zhuying Wang[1,2#], Huachen Rao[3,#], Yulei Han[4,#], Bo Liu[1], Shuikang Yu[1], Yunmei Zhang[1], Ruiqing Zang[1], Runqing Luan[1], Shuting Peng[1], Zhenhua Qiao[1,2,3], Zhenyu Wang[1,2,*], Junfeng He[1,*], Tao Wu[1,2,3,5*] and Xianhui Chen[1,2,3,5*]

[1]Department of Physics, University of Science and Technology of China, Hefei, Anhui 230026, China.

[2]Hefei National Laboratory, University of Science and Technology of China, Hefei 230088, China.

[3]Hefei National Research Center for Physical Sciences at the Microscale, University of Science and Technology of China, Hefei, China.

[4]Department of Physics, Fuzhou University, Fuzhou, Fujian 350108, China.

[5]Collaborative Innovation Center of Advanced Microstructures, Nanjing University, Nanjing 210093, China.

[#]These authors contributed equally to this work.

*To whom correspondence should be addressed: zywang2@ustc.edu.cn; jfhe@ustc.edu.cn; wutao@ustc.edu.cn; chenxh@ustc.edu.cn

A central mystery in high-temperature cuprate superconductors is the coexistence of multiple exotic orders, which is presumably associated with strong electronic correlation. The ongoing interest in this enigmatic phenomenon is further energized when similar electronic orders and states emerge and coexist in less correlated Kagome superconductors. Here, by utilizing angle-resolved photoemission spectroscopy (ARPES), nuclear magnetic resonance (NMR) spectroscopy, scanning tunneling microscopy (STM) and first-principles calculations, we reveal the sudden emergence of a distinct short-range charge stripe order in Sn-doped $CsV_3Sb_5$ Kagome superconductors when the long-range 2 x 2 charge density wave order in pristine $CsV_3Sb_5$ is suppressed. This short-range stripe order features a modulation vector of approximately 1/3 along one of the three lattice directions, induces remarkable quasiparticle scattering between the original quasi-1D Kagome-*d*-bands and their replica of folding, and clearly suppresses the electron density of states at the Fermi level. Our first-principles calculations reveal that 3 × 1 supermodulation represents a hidden secondary instability in pristine $CsV_3Sb_5$. This instability is further enhanced by in-plane chemical pressure induced by Sn substitution, and coupled to the electronic correlation, leading to a unique charge stripe order in the system. As **such, our results reveal a new route toward emergent electronic orders via cooperative interactions between the lattice and electronic degrees of freedom.**

## I. INTRODUCTION

In Landau theory, different phases in materials are characterized by different broken symmetries, and the transition between two phases is associated with a sudden reduction in energy. However, in correlated quantum materials, most notably high-temperature superconductors, multiple ordered states with different broken symmetries are nearly degenerate in energy [1-3]. As such, different orders may coexist over a large range of the phase diagram [4]. These orders were recently described as intertwined orders [5], but their underlying driving mechanisms are still elusive. Therefore, realizing intertwined orders in other quantum materials and revealing the distinct behaviors of the delicate transitions between these nearly degenerate states are highly desirable at the current stage.

Kagome metals $AV_3Sb_5$ (A = K, Rb, and Cs) have attracted intense interest because of the emergence of many exotic orders and states [6-52,55,56], ranging from superconductivity [6-10], charge density waves (CDWs) [6,7,11-20], nematic order [21-23], pair density waves [24], topologically nontrivial states [25,26], one-dimensional stripe order [27-29] and time-reversal symmetry breaking states [30-32]. In particular, increasing experimental evidence and theoretical analysis suggest a possible intertwined nature between these emergent orders [6,10,15,21,24,28,32,33]. Compared with high-temperature cuprate superconductors, where strong electron correlation dominates, $AV_3Sb_5$ Kagome superconductors are less correlated in the context of onsite Coulomb repulsion. Nevertheless, interactions from multiple channels have been proposed in Kagome superconductors, including lattice instabilities [12,34-38], Fermi surface instabilities [12,38-41], effective electron correlation associated with van Hove singularities [42-45], and sublattice interference with intersite interactions [33,46]. Therefore, Kagome superconductors provide a new platform for understanding the diversity of intertwined orders and the associated driving mechanism.

## II. RESULTS

We have investigated the Sn-doped Kagome superconductor $CsV_3Sb_{5-x}Sn_x$, in which the substitution of Sb by Sn atoms presumably induces minimal changes to the V-based Kagome sublattice [29]. Remarkably, the doping-dependent phase diagram of $CsV_3Sb_{5-x}Sn_x$ exhibits two complete superconducting domes, mimicking the phase diagram of $CsV_3Sb_5$ under pressure, where unconventional density wave orders and superconductivity are reported [28,47]. Transport and specific heat capacity measurements indicate that the long-range 2 × 2 CDW order in pristine $CsV_3Sb_5$ suddenly disappears in the $CsV_3Sb_{5-x}Sn_x$ compound at a critical doping level of x ~ 0.07 (Figs. 1a-d; also see Fig. S1). This sudden change in the electronic structure is also revealed by ARPES measurements (Figs. 1i-m). As shown in Fig. 1i, the 2 × 2 CDW order in pristine $CsV_3Sb_5$ leads to a characteristic double-band feature around the M point at ~0.6 eV below $E_F$ (marked in red) [38,48]. This feature remains discernible with Sn doping at x ~ 0.06, but it changes to a single-band feature at x ~ 0.07, indicating the absence of the 2 × 2 CDW order. We define the energy difference between the two bands at the M point as $\Delta$ (marked in Fig. 1i) and plot its magnitude as a function of Sn doping (Fig. 1m). An abrupt change is clearly shown at x ~ 0.07, indicating the sudden collapse of the 2 × 2 CDW order. This sudden change in CDW order is further confirmed by bulk $^{51}$V NMR measurements (see Appendix A for experimental details). As shown in Fig. 1f, the 2 × 2 CDW order in pristine $CsV_3Sb_5$ results in a characteristic double-peak structure in the central NMR transition line of $^{51}$V nuclei at 2 K, which is ascribed to triangle and hexagonal clusters in the so-called inverse Star-of-David pattern (Fig. 1h) [49]. With increasing the Sn doping level to x ~ 0.07, the double-peak structure suddenly changes into a single NMR peak, whereas the average Knight shift ($\overline{K}$) of the overall central transition lines experiences a sudden jump (Fig. 1e). All these results support a sudden collapse of the 2 × 2 CDW order at x ~ 0.07. Furthermore, by comparing the NMR line shape at x ~ 0.07 with that of pristine $CsV_3Sb_5$, we find that it becomes significantly broadened and asymmetric, strongly suggesting possible short-range CDW order. Above x ~ 0.07, in addition to a slight change in asymmetry, the overall NMR spectra at 2 K do not exhibit a significant change in linewidth with increasing Sn doping level. Interestingly, the superconducting transition temperature also shows a sudden suppression at x ~ 0.07 and then gradually increases with increasing Sn doping (Fig. 1b). This result indicates that the short-range CDW order is not just a residual 2 × 2 CDW order. In pressurized $CsV_3Sb_5$, a similar sudden suppression of superconductivity originates from the emergence of a commensurate stripe-like CDW order [28]. Notably, the pressure-induced stripe-like CDW order also leads to a sudden jump in the average Knight shift, similar to that in our case (Fig. S3). Moreover, the overall NMR central transition line for x ~ 0.07 is highly similar to that of the pressure-induced stripe-like CDW order (Fig. S4). In this sense, stripe-like CDW offers a natural explanation for the observed short-range CDW order and the sudden change in the superconducting transition temperature at x ~ 0.07.

To establish a microscopic understanding of the new charge-stripe order, we have carried out a detailed scanning tunneling microscopy study at 4.2 K for the CsV$_3$Sb$_{5-x}$Sn$_x$ series. The Sn dopants appear as dark depressions in the topographies, with their centers located right in the middle of the six topmost Sb atoms (Fig. 2b), indicating that the Sn dopants favor substituting Sb atoms in the V$_3$Sb Kagome plane. Consistent with the APRES results, from the topographic images, we find that the parent 2 × 2 CDW order suddenly disappears when the local Sn concentration exceeds a critical value of x ~ 0.06 (Fig. S5). However, the presence of surface singe-*Q* 4a$_0$ order [27] from x = 0 to 0.09 hinders the confirmation of the new charge stripe from topographies. Therefore, we turn to quasiparticle interference (QPI) imaging, with which the reconstruction of band dispersion due to emergent bulk stripe order, if any, can be revealed.

Figures 2e-h show the Fourier transforms of typical differential conductance maps at energies near the Fermi level for x = 0, 0.06, 0.07 and 0.09 that span the critical doping level (see Appendix A). Once again, we do not find a notable difference between the QPI patterns for x=0 and 0.06, which exhibit the characteristic patterns of a 2 × 2 CDW order: The main QPI channels arise from intra-*p*-orbital scattering ($q_0$), scattering of the quasi-1D segments of the Kagome-*d*-orbital ($q_1$ and $q_3$), and scattering between the arcs of triangular pockets ($q_2$) [27,50], which are schematically shown in Fig. 2k. Here, we emphasize that the extracted energy dispersion of these bands perfectly matches the DFT calculations based on a staggered inverse star-of-David CDW structure [50], suggesting that the surface 1Q-4a$_0$ order does not dramatically reconstruct the band structure. In contrast, for samples with x = 0.07 and 0.09 in which the bulk charge-stripe order starts to develop, we find two new, long quasi-1D stripes in the QPI pattern (labeled as q' and q'' in Fig. 2g-i) that span almost the entire Brillouin zone. The q-space positions of these two signals disperse noticeably in the energy window we measured (Fig. 2j), demonstrating that they are not simply formed by the real-space patterns resulting from the impurity locations but rather by the QPI signatures that are related to the band structure (Figs. S6-7). The observation of these quasi-1D features in the QPI strongly suggests that new scattering channels arise from some parallel quasi-1D segments in the constant energy contours (CECs) of the charge-stripe-ordered samples.

To explore the origin of these new scattering channels, we first check their energy dispersion. By extracting the dispersion along the orange arrows shown in Figs. 2i-j, we find that their slope is comparable to that of the quasi-1D segments of the Kagome-*d*-orbital: the extracted dispersion from QPI patterns is approximately 4.3 eV·Å, which well matches the value of the quasi-1D vanadium d$_{xz/yz}$ orbitals (approximately 4 eV·Å) obtained from the ARPES measurements [51]. Since the magnitudes of q' and q'' along Γ-K are approximately 1/3 and 2/3 of $\sqrt{3}/2Q_{Bragg}$ (corresponding to the middle point between two neighboring Bragg peaks), it is reasonable to assume that the emergence of charge-stripe order in x = 0.07 and 0.09 folds the quasi-1D Kagome-*d* bands with a vector of ~1/3 with respect to one of the three lattice directions, and new channels are raised from the scattering between their original and replica bands, as shown in Fig. 2l. This assignment requests a 1Q-3 × 1 order that can give rise to the observed QPI data (see Appendix A). To test this hypothesis, we have performed measurements on a x = 0.09 sample by applying 0.5% uniaxial tensile strain along one lattice direction, as this process can significantly suppress the surface $4a_0$ reconstruction. As shown in Fig. 2m, while the surface $4a_0$ order becomes almost invisible under strain (Fig. S8), the QPI signals still exhibit the quasi-1D features of q' and q'' that are almost identical to those of the unstrained sample. This finding further suggests that these two features originate from the emergent charge stripe in the bulk; more importantly, a short-range 1Q-3 × 1 order can now be directly observed in the differential conductance maps (Figs. 2m and n). The orientation of this order rotates 120 degrees with respect to the weak, remanent local surface $4a_0$ modulation. Therefore, taking the QPI analysis and real-space mapping together, our data strongly suggest that a new 1Q -3 × 1 stripe order emerges in the samples with x = 0.07 and 0.09. We note that previous X-ray diffraction (XRD) experiments reported an incommensurate, short-range stripe-like CDW with $Q \approx 0.37$ in a heavily Sn-doped $CsV_3Sb_5$ (x = 0.15) [29] and a long-range stripe order with $Q \approx 3/8$ in a pristine sample under pressure [52]. These *Q*-vectors are comparable to our findings but slightly different. Owing to the short-range nature of the charge-stripe order and different sensitivities to the 1D short-range order of STM and XRD [53,54], it is reasonable to conclude that the charge stripe order observed in these systems may share similar physics, and future work exploring the doping dependence of this charge ordering is merited.

Next, to shed light on the underlying physics, we further investigate the electronic states near the Fermi level in this short-range stripe order regime. As shown in Figs. 3a-f, we systematically studied the temperature-dependent evolution of the nuclear spin-lattice relaxation rate ($1/T_1$) in the $CsV_3Sb_{5-x}Sn_x$ samples. In a Fermi liquid, the nuclear spin-lattice relaxation rate divided by the temperature ($1/T_1T$) is proportional to $N(E_F)^2$, where $N(E_F)$ represents the total density of states at the Fermi level $E_F$. In pristine $CsV_3Sb_5$, besides a sudden jump at $T_{CDW}$, the temperature-dependent $1/T_1T$ shows moderate suppression below $T_{CDW}$, indicating a partial gap in a 2 × 2 CDW state. This is consistent with previous ARPES observations of the momentum-dependent CDW gap [38,55]. With increasing Sn doping, although the long-range 2 x 2 CDW order suddenly disappears at x ~ 0.07, the temperature-dependent $1/T_1T$ always shows a continuous suppression at low temperatures (Fig. S10). This fact indicates that the short-range stripe order still opens a partial gap at the Fermi level. To quantify the reduced density of states at the Fermi level, we calculate the quantity of $(T_1T(300K)/T_1T(10K))^{1/2}$, which is proportional to $N(E_F,10K)/N(E_F,300K)$. As shown in Fig. 3g, $N(E_F,10K)/N(E_F,300K)$ exhibits a sudden suppression when the short-range stripe order appears at x ~ 0.07. This result strongly suggests that, compared with the long-range 2 × 2 CDW order, the formation of short-range stripe order results in a more depleted density of states at the Fermi level. The suppression of the electron density of states in the short-range stripe order regime observed by NMR is also confirmed by ARPES measurements (Fig. 3h-n). As shown in Figs. 3i-n, measurements of the sample with x ~ 0.09 clearly reveal a suppression of the electron density of state near $E_F$ in the band dispersion along Γ−K−M (Fig. S11) at a low temperature (10 K, Figs. 3i-k), which disappears at an elevated temperature (90 K, Figs. 3l-n). To visualize the electron density of states suppression, photoemission spectra between the K and M points are integrated and shown as a function of temperature (Fig. 3h). A gradual suppression of the spectral weight is clearly identified at the Fermi level while the temperature decreases (Fig. 3h). While the density of state suppression observed by both NMR and ARPES is naturally associated with the short-range stripe order (Fig. S12), it remains important to consider the possibility of a remanent 2 × 2 CDW order, which cannot be probed by transport and specific heat capacity measurements. First, earlier studies revealed that the 2 × 2 CDW gap opening in doped $CsV_3Sb_5$ (e.g., Ti-doped $CsV_3Sb_5$) disappears immediately in the ARPES spectrum when the signature of the 2 × 2 CDW order disappears in the transport measurement (Fig. S13). Second, to check any possible local 2 × 2 CDW order, STM measurements have been performed on the same piece of sample ($CsV_3Sb_{5-x}Sn_x$, x = 0.09) immediately after the ARPES measurements. No indication of a local 2 × 2 CDW order is observed (inset of Fig. 3h). Third, any possible density of state suppression by the remanent 2 × 2 CDW order cannot produce the nonmonotonic doping-dependent behavior observed via NMR (Fig. 3g). As such, the totality of the experimental results points to an intrinsic suppression of the electron density of states due to the short-range stripe order.

## III. DISCUSSION AND CONCLUSION

We note that the doping-dependent evolution of the CDW orders is counterintuitive to empirical expectations. Typically, when the long-range 2 × 2 CDW order is suppressed with doping, one would expect either a short-range 2 × 2 CDW order that persists over a certain doping range or a sharp phase transition that leads to the formation of another long-range order. However, our measurements reveal the sudden emergence of a distinct short-range stripe order. In the following, we discuss the possible origin of this short-range stripe order.

First-principles calculations have been carried out to determine the total energy of pristine $CsV_3Sb_5$ with different lattice structures (Figs. 4a-b). Similar to earlier reports [12,56], we find that either the Star of David (SD) structure or the Inverse Star of David (ISD) structure exhibits a total energy lower than that of the Kagome lattice (Fig. 4a), leading to a natural instability toward the 2 × 2 CDW order. Strikingly, our calculations reveal two other structures that are also energetically more stable than the pristine Kagome lattice in $CsV_3Sb_5$ (Fig. 4b, labeled as type 1 and type 2, respectively). These two structures represent a 3 × 1 supermodulation of the original Kagome lattice, which coincides with the wavevector of the short-range stripe order revealed by our STM measurements. In this regard, the 3 × 1 supermodulation represents a hidden secondary instability of the pristine $CsV_3Sb_5$ compound.

The next step is to understand the role of Sn substitution in promoting the 3 × 1 supermodulation to a leading instability in the Sn-doped $CsV_3Sb_5$. From STM imaging, we find that all the Sn dopants substitute Sb atoms in the $V_3Sb$ Kagome plane, therefore having minimal influence to the vanadium Kagome network. This atomic position of dopants could be a key factor for the emergence of the charge stripe, as any substitution to the vanadium Kagome network (e.g., Ta, Ti and Nb), thus far, does not induce such an order, even though that both Ti and Sn substitutions dope holes into the system. Such type of Sn substitution would directly change the lattice chemical pressures, leading to an extension of the in-plane lattice constant *a* and thus a reduction in the lattice ratio of *c/a* [57]. In this context, we have performed first-principles calculations to qualitatively simulate the effect of in-plane pressure. For simplicity, we gradually decrease the lattice ratio *c*/a and calculate the phonon spectrum of $CsV_3Sb_5$. Surprisingly, the minimum of the negative phonon frequency shifts from M (L) to a position at ~2/3 Γ-M (A-L) (see Fig. 4c and Fig. S14), indicating that the wavevector of the lattice instability changes from 1/2 to ~1/3. We note that these two wavevectors match those of 2 × 2 CDW modulation and 3 × 1 stripe order, respectively. The lattice instability finally disappears with further reduction of the lattice ratio *c*/a (Fig. 4c), which is also consistent with the absence of any charge order in samples with excessive Sn substitutions. We note that these calculations are simplified to mainly consider the effect of in-plane pressure, other detailed changes induced by Sn substitution are ignored. In this context, it is encouraging that such a simple simulation qualitatively captures the experimental observations. These results indicate that the in-plane lattice chemical pressure induced by Sn is indeed helpful in driving the transition from 2 × 2 CDW modulation to the new 3 × 1 stripe order.

While the above calculations have substantially deepened our understanding of the experimentally observed 3 × 1 stripe order, we note that this new order also exhibits properties beyond the first-principles calculations. For comparison, electron energy bands are calculated for the 2 × 2 ISD (Fig. 4d), Kagome (Fig. 4e) and stripe (type 1, Fig. 4f) structures. A partial gap opening is observed at the Fermi level for the 2 × 2 ISD structure (Fig. 4d), similar to the experimental observation. However, the calculated electronic energy bands for the 3 × 1 stripe structure exhibit little change from those of the Kagome lattice near the Fermi level (Figs. 4e-f). Sn doping induces clear band reconstructions at high binding energies for the 3 × 1 stripe structure, but no gap opening is shown near the Fermi level (Fig. 4g). This is distinct from the experiments, where a partial gap opening and a suppression of the electronic density of states are identified by multiple techniques. In this context, electronic correlation effects beyond density functional theory (DFT) are required to account for the gap opening in the experiments.

Usually, strong electronic correlation can renormalize the band structure effectively. Notably, this is obviously not the case for CsV$_3$Sb$_5$. Theoretically, the onsite electronic correlation is strongly suppressed in a 2D Kagome lattice owing to the sublattice interference effect, but nonlocal electronic correlations, on the other hand, are promoted near the van Hove filling [46]. However, determining such a nonlocal electronic correlation effect in the Kagome lattice remains elusive and challenging at present stage. The rich symmetry breaking in the parent 2 × 2 CDW state is widely believed to be a strong indication of such a nonlocal electronic correlation, especially for the putative time-reversal symmetry breaking, which cannot be understood without electronic correlations. In the current case, the Sn substitution might also play a critical role on electronic correlations related to the van Hove singularities (vHSs) near E$_F$. With increasing Sn doping, the bottom of the electron-like Sb-p pocket at the Γ point shifts dramatically closer to the Fermi level with hole doping, the vHSs at the M point, however, exhibit little change (Fig. 1i-m and Fig. S15), sustaining the non-local Coulomb interaction promoted by these vHSs. The resulted band structure is in good agreement with the calculations for pressurized pristine CsV$_3$Sb$_5$ in which a long-range stripe order has been revealed [52,58,59]. In pressurized CsV$_3$Sb$_5$, a similar stripe-like CDW appears at pressures P > 0.6 GPa, and is completely suppressed above 2 GPa [28,52]. Previous nuclear spin-lattice relaxation measurement has revealed that, by suppressing the stripe-like CDW, the temperature-dependent 1/$T_1T$ exhibits a significant increase beyond quasiparticles contribution at low temperatures, indicating remarkable spin fluctuations [28]. Enhanced spin fluctuations could serve as strong evidence for the underlying electronic correlations in pressurized $CsV_3Sb_5$ while a complete microscopic understanding is still lacking theoretically. In Sn-doped samples, when the short-range stripe order develops at low temperatures for x > 0.06, the $1/T_1T$ results clearly indicate that the above-mentioned spin fluctuations are strongly suppressed (Fig. S16), suggesting a release of the electronic correlations by forming the charge-stripe order. In addition, compared to the triple-$q$ CDW order in pristine $CsV_3Sb_5$, the unidirectionality of the charge stripe in Sn-doped samples, i. e., the breaking of rotational symmetry, is not predicted from the calculations and requests additional interaction from the electronic degree of freedom. Considering all these facts, electronic correlations must play an assistant role to drive the stripe-like CDW.

Finally, the absence of long-range stipe order remains to be understood. Possible scenarios include strong competition between the 2 × 2 CDW and the 3 × 1 stripe order, as well as the disorder effects induced by Sn doping. Further theoretical efforts are also needed to investigate the microscopic mechanism for the cooperation between lattice instability and electronic correlation in the context of intertwined orders. Our present finding of electron-correlation-assisted charge stripe order also suggests an enhanced interplay between superconductivity and charge order above x ~ 0.06 in the Sn-doped $CsV_3Sb_5$ system. In a previous report, the suppression of superconductivity with moderate Sn doping was attributed to a change in the three-dimensional stacking of the 2 × 2 CDW [60], i.e., from the LLL + MMM (SoD + TrH) phase in pristine sample to MLL (staggered ISD) phase in the doped sample. With the discovered charge-stripe order and the updated phase diagram, we can conclusively link the suppression of superconductivity to the emergence of charge-stripe order that occur concurrently at x ~ 0.07. Here, similar electronic correlations might affect the charge stripe order and superconductivity simultaneously, which may lead to a possible intertwinement between the charge stripe order and superconductivity. Further investigation of the superconducting state above x ~ 0.06 is urgently needed to explore possible intertwined electronic orders.

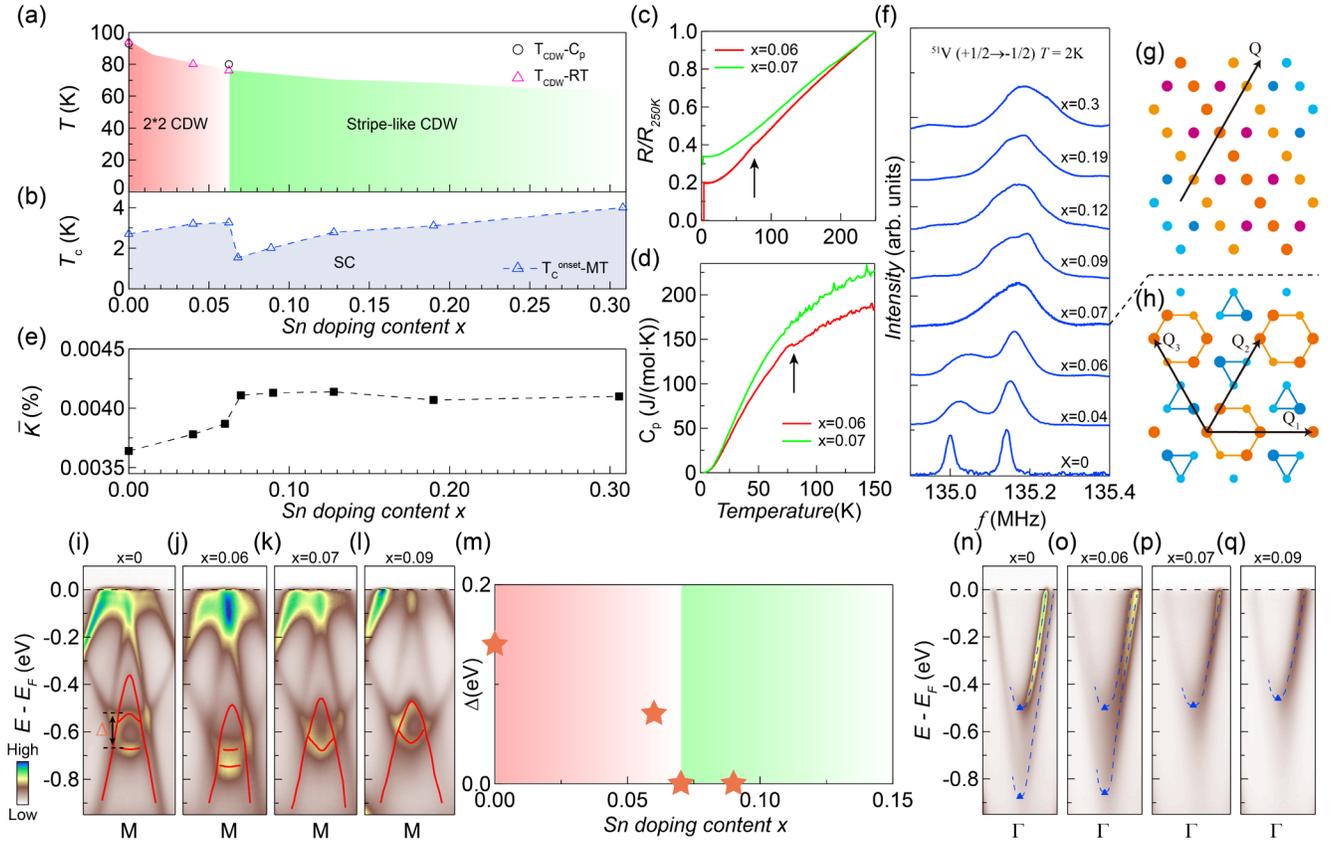

**FIG. 1**. Sudden suppression of the 2 × 2 CDW order in Sn-doped $CsV_3Sb_5$. (a), Doping-dependent phase diagram. The CDW transition temperature for 2 × 2 CDW is defined by the specific heat capacity (black circles) and *ab*-plane resistivity (pink triangles). The green area represents a schematic region for the short-range stripe-like CDW. (b), Doping-dependent superconducting phase diagram. The superconducting transition temperature ($T_c$) is determined by magnetic susceptibility measurements (Fig. S1). The sudden suppression of $T_c$ at x = 0.07 is associated with the appearance of short-range stripe-like CDW. (c), Temperature-dependent *ab*-plane resistivity for $CsV_3Sb_{5-x}Sn_x$ with x = 0.06 and 0.07. (d), Temperature-dependent specific heat capacity for $CsV_3Sb_{5-x}Sn_x$ with x = 0.06 and 0.07. (e), Doping-dependent evolution of the average Knight shift ($\bar{K}$) extracted from the $^{51}$V NMR central transition lines, where $\bar{K} = \frac{\int_f (I(f) \times f) df / \int_f I(f) df}{H \times \gamma_N} - 1$. (f), $^{51}$V NMR central transition lines at 2 K in different $CsV_3Sb_{5-x}Sn_x$ samples. The external magnetic field of $\mu_0 H$ =12 T is applied along the crystalline *c* axis. The temperature-dependent evolution of the $^{51}$V NMR central transition lines for all the samples is shown in the supplementary materials (Fig. S2). (g), (h), Illustrations of the doping-dependent evolution of the V sublattice, corresponding to the stripe-like superlattice (g) and the ISD superlattice (h), respectively. The color circles represent the lattice symmetry in the V sublattice. (i)-(l), Photoelectron intensity plot of the band structure around M point (along Γ-K–M direction) of $CsV_3Sb_{5-x}Sn_x$ samples with x = 0 (i), 0.06 (j), 0.07 (k) and 0.09 (l), measured with 56 eV photons at 25 K (7 K) for the pristine (doped) samples. The red solid line is a guide to the eye. Δ is defined in (i) to characterize the energy difference between the two features of the 2 × 2 CDW. (m), Doping evolution of Δ. (n-q), Doping evolution of the electron-like band around Γ point, measured on $CsV_3Sb_{5-x}Sn_x$ samples with x = 0 (n), 0.06 (o), 0.07 (p) and 0.09 (q), respectively.

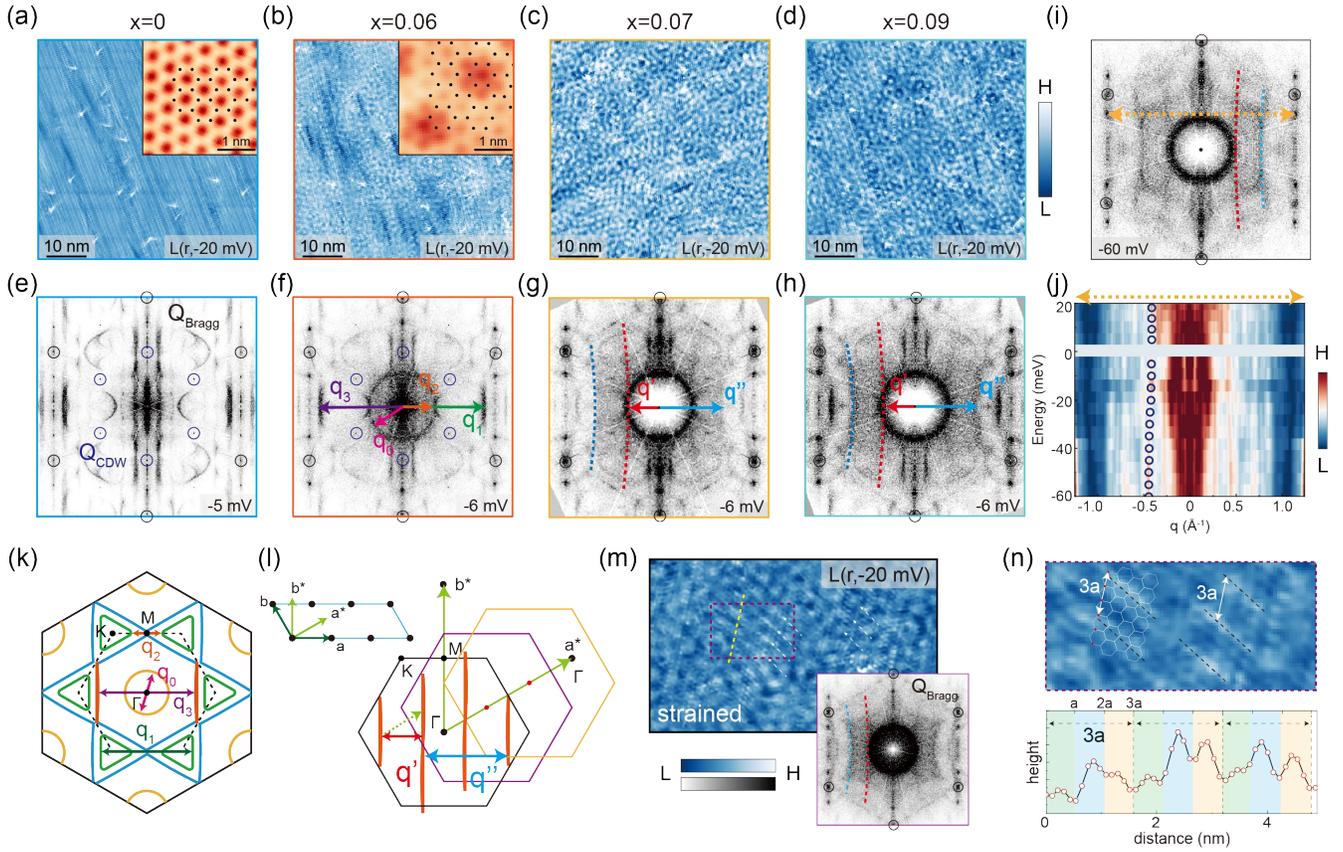

**FIG. 2.** Charge–stripe order with a wavevector of approximately 1/3 for x=0.07 and 0.09. (a)-(d), Typical differential conductance $L$(r, −20 meV) maps of the Sb-terminated surface for x = 0, x = 0.06, x = 0.07 and x = 0.09. The insets in (a) and (b) show the Sb honeycomb lattice (marked by dark dots) and two Sn dopants, respectively. (e)-(h), Fourier transforms (FTs) of differential conductance maps near the Fermi level. Characteristic scattering wavevectors $q_i$ are denoted by color-coded arrows in (f) and (k). The Bragg and 2×2 CDW peaks are marked with black and purple circles, respectively. New scattering channels are labeled as q' and q'' in (g) and (h) for x= 0.07 and 0.09. (i), QPI pattern for x=0.09 but at a high energy of -60 meV. (j), Energy dispersion of the q' vector extracted from the QPI along the orange line in (i). (k), Schematic constant energy contours (CECs) near the Fermi level for x=0. (l), Illustration of the original lattice vectors (green arrows) and new vectors of 1Q-3 × 1 order (bule) in real space. The lower panel shows how the quasi-1D bands, in the reciprocal space, are folded for the 1Q-3 x 1 order along ΓM (dashed green arrow), which generates a new scattering channel along ΓK. The colored arrows q' and q'' denote these new scattering wave vectors. (m), Differential conductance map $L$(r, -20 meV) for a strained sample (x=0.09) with invisible surface $4a_0$ order. The FT clearly shows similar scattering signatures of q' and q'' that are almost identical to the unstained sample. (n), Zoomed-in view for the region marked by purple dashed rectangle in (m), highlighting the existence of a short-range 3 × 1 order. The lower panel shows a line profile of the differential conductance taken along the yellow dashed line in (m), with which a modulation of $3a_0$ can be directly observed.

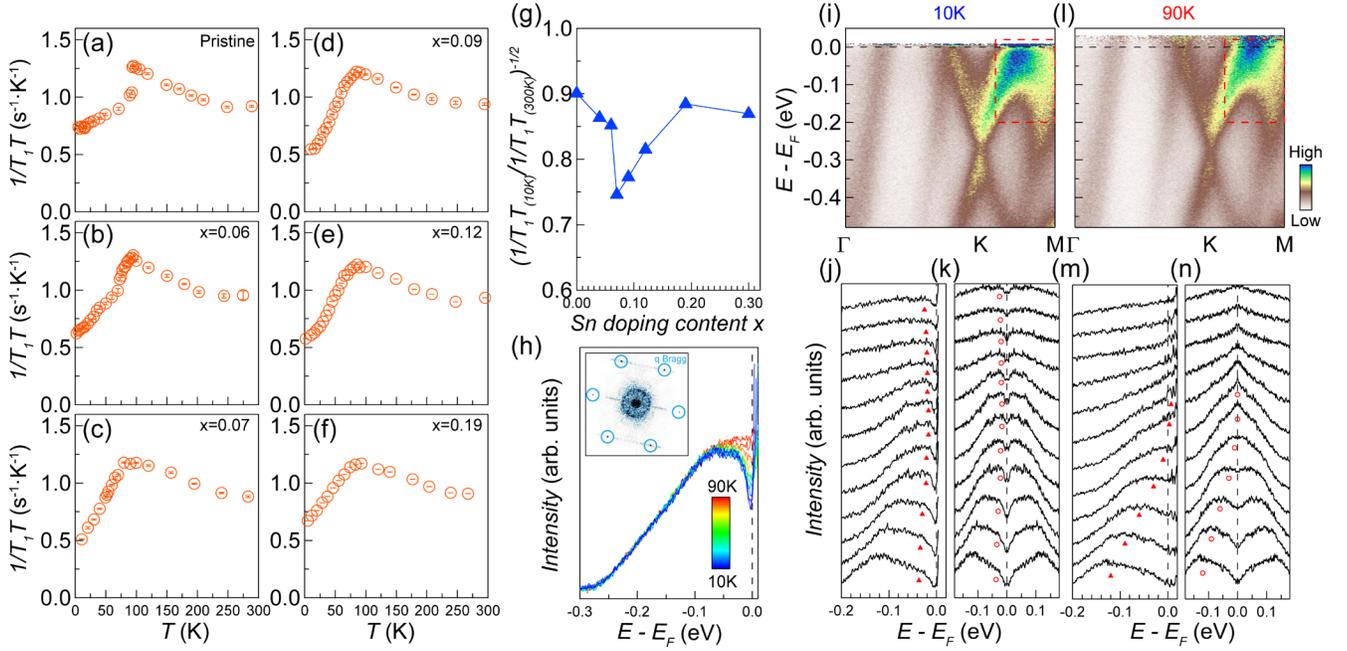

**FIG. 3.** Suppression of the electron density of states at the Fermi level in Sn-doped $CsV_3Sb_5$. (a)-(f), Temperature-dependent $1/T_1T$ in different $CsV_3Sb_{5-x}Sn_x$ samples with x = 0 (a), 0.06 (b), 0.07 (c), 0.09 (d), 0.12 (e) and 0.19 (f). For x ≤ 0.06, the low-temperature $1/T_1T$ below the $T_{CDW}$ is calculated by averaging the $1/T_1T$ measured on both splitting peaks (Fig. S9). The high-temperature $1/T_1T$ is measured at the top of a single peak. For x ≥ 0.07, the $1/T_1T$ is always measured at the top of the broad peaks. (g), The quantity of $(T_1T(300K)/T_1T(10K))^{1/2}$ as a function of Sn doping. (h), Integrated ARPES spectra between the K and M points as a function of temperature measured for the $CsV_3Sb_{5-x}Sn_x$ (x = 0.09) sample. (i), Photoelectron intensity plot of the band structure along Γ-K-M measured at 10 K. (j)-(k), EDCs in the red dotted boxes in (i). The EDCs are divided by the Fermi–Dirac function (j) and symmetrized with respect to $E_F$ (k). The EDC peaks are marked by triangles and circles. (l)-(n), Same as (i)-(k) but measured at 90 K.

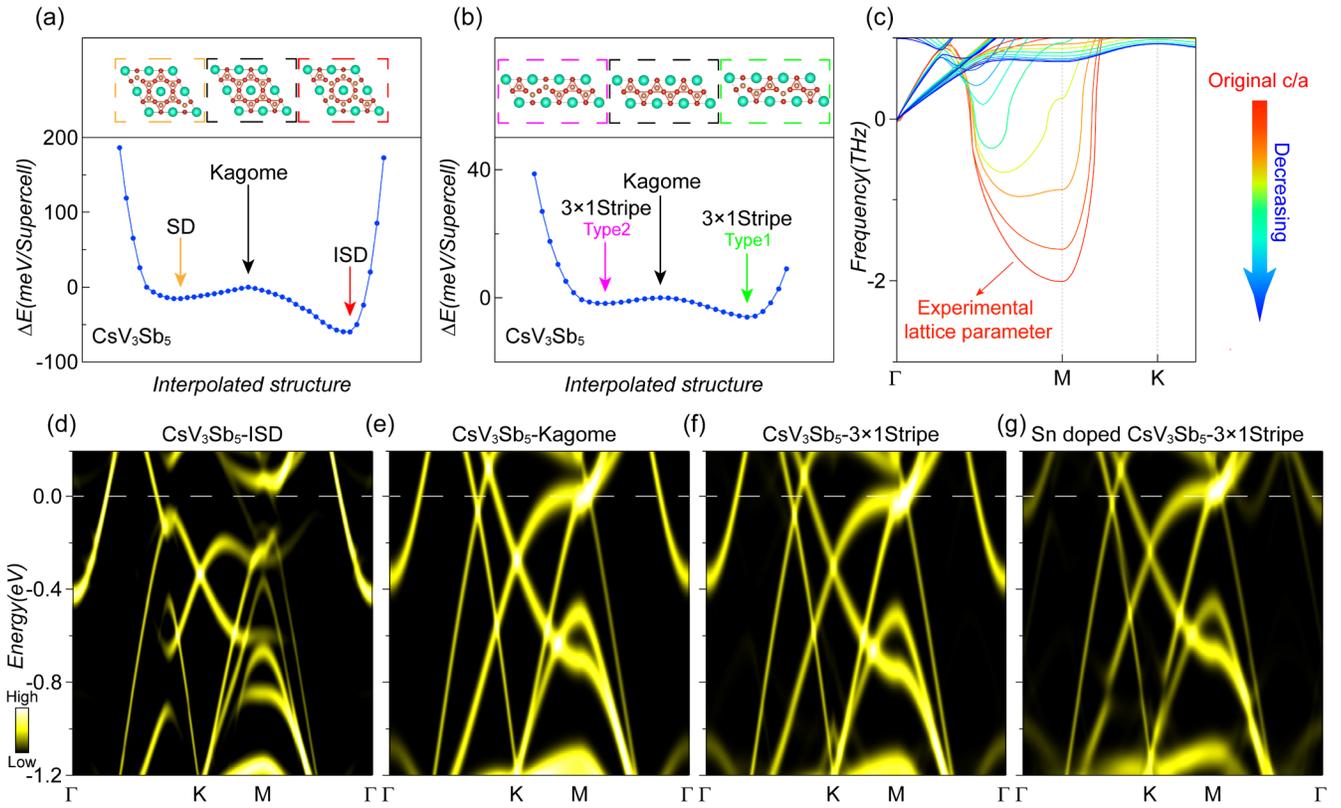

**FIG. 4.** Stripe instability in pristine and Sn-doped CsV$_3$Sb$_5$. (a), Calculated total energy of pristine CsV$_3$Sb$_5$ as a function of the interpolated structure toward the SD and ISD structures. ΔE represents the relative total energy with respect to that of the Kagome structure per supercell (36 atoms). The 2 × 2 supercells for the Kagome structure, SD structure and ISD structure are shown in the inset. (b), Same as (a) but shows the calculated total energy as a function of the interpolated structure toward two types of 3 × 1 stripe structures. ΔE is also calculated in the 3 × 1 supercell (27 atoms). (c), Phonon spectrum of CsV$_3$Sb$_5$ near M point as a function of the lattice ratio c/a (from 100% to 90% of the experimental value). (d)-(g), Calculated band structure along Γ-K–M-Γ for CsV$_3$Sb$_5$ with an ISD structure (d), CsV$_3$Sb$_5$ with a Kagome structure (e), CsV$_3$Sb$_5$ with a 3 × 1 Stripe structure (f) and Sn-doped CsV$_3$Sb$_5$ with a 3 × 1 Stripe structure (g). (h), Schematic energy changes with electron correlation.

**Acknowledgments**

The work at the University of Science and Technology of China (USTC) was supported by the National Key R&D Program of the MOST of China (Grant No. 2022YFA1602601, 2024YFA1408100), the National Natural Science Foundation of China (Nos. 12488201, 52273309, 52261135638, 12034004, 12161160316, 12325403, 12474158 and 12488101), the Chinese Academy of Sciences under contract No. JZHKYPT-2021-08, the CAS Project for Young Scientists in Basic Research (Grant No. YBR-048), the Innovation Program for Quantum Science and Technology (Grant No. 2021ZD0302800), the Fundamental Research Funds for the Central Universities (Nos. WK3510000015, WK3510000012, WK9990000110), the International Partnership Program of the Chinese Academy of Sciences (No. 123GJHZ2022035MI) and the Innovation Program for Quantum Science and Technology (No. 2021ZD0302802). The use of the Stanford Synchrotron Radiation Lightsource, SLAC National Accelerator Laboratory, is supported by the U.S. Department of Energy, Office of Science, Office of Basic Energy Sciences under Contract No. DE-AC02-76SF00515. We are grateful to the Supercomputing Center of the University of Science and Technology of China for providing high-performance computing resources.

## APPENDIX A: MATERIALS AND METHODS

### 1. Sample preparation and characterization

Single crystals of $CsV_3Sb_{5-x}Sn_x$ were grown by using $Cs_2Sb_3$, $VSb_2$ and $SnSb$ alloys as fluxes. For the growth of different Sn doping levels, the ratio of $Cs_2Sb_3$ to $VSb_2$ was fixed at 2:3. The mixtures of the three precursors were placed in an $Al_2O_3$ crucible and then sealed in a quartz tube. The above procedure was performed in a glove box with an argon atmosphere. The sealed tubes were first heated to 1273 K and then maintained at this temperature for 24 hours. Finally, single-crystal growth was achieved by slow cooling to 773 K. Single crystals were picked up from the flux by mechanical exfoliation. The chemical composition was determined via energy-dispersive X-ray spectroscopy (GeminiSEM, 500). Electrical transport and specific heat capacity were measured via a Quantum Design Physical Property Measurement System (PPMS). Magnetic susceptibility was measured with a SQUID magnetometer (Quantum Design MPMS-5) and a magnetic property measurement system (Quantum Design SQUID-VSM) with the $^3$He option.

## 2. High-resolution ARPES measurements

High-resolution angle-resolved photoemission measurements were carried out at Beamline 5-2 of the Stanford Synchrotron Radiation Lightsource (SSRL) of the SLAC National Accelerator Laboratory using 56 eV photons with a total energy resolution of ~10 meV and at our laboratory-based ARPES system using 21.2 eV photons with a total energy resolution of ~3 meV. The Fermi level was referenced to that of polycrystalline Au in electrical contact with the samples. The base pressure for all the measurements was better than $5 \times 10^{-11}$ torr.

## 3. NMR measurements

A commercial NMR spectrometer from Thamway Co. Ltd. was used for the NMR measurements. An NMR-quality magnet from Oxford Instruments offers a uniform magnetic field of up to 12 Tesla. The external magnetic field at the sample position was calibrated by $^{63}$Cu NMR of the NMR coil. All NMR spectra were measured via the standard spin echo method with a fast Fourier transform (FFT) sum. All NMR measurements were performed under an external magnetic field of 12 T along the *c* axis. For $^{51}$V nuclei, the nuclear spin number is *I* = 7/2, and the gyromagnetic ratio $\gamma_N$ is 11.193 MHz/T. Since the NMR frequency of $^{51}$V nuclei is strongly dependent on its surrounding structural environment, the doping of tin atoms produces two additional NMR peaks, which are ascribed to the V site, with one tin atom at the nearest neighbor in-plane Sb site and two tin atoms at the nearest neighbor in-plane Sb site (Fig. S17). The nuclear spin-lattice relaxation time ($T_1$) of $^{51}$V nuclei was measured via the inverse recovery method, and the recovery of the nuclear magnetization M(t) was fitted by a function with $1 - \frac{M(t)}{M(\infty)} = I_0 \left( \frac{1}{84} \times \exp\left(-\left(\frac{t}{T_1}\right)^\beta\right) + \frac{3}{44} \times \exp\left(-\left(\frac{6t}{T_1}\right)^\beta\right) + \frac{75}{364} \times \exp\left(-\left(\frac{15t}{T_1}\right)^\beta\right) + \frac{1225}{1716} \times \exp\left(-\left(\frac{28t}{T_1}\right)^\beta\right) \right)$, in which error bars were determined via the least-square method.

## 4. First-principles calculations

First-principles calculations were performed via the projected augmented-wave method [61] as implemented in the Vienna ab initio simulation package (VASP) [62,63]. The exchange–correlation interaction was addressed via the Perdew–Burke–Ernzerhof type of generalized gradient approximation [64]. The cutoff energy for the plane wave basis was set to 500 eV. The thresholds for energy convergence and Hellmann–Feynman forces were set to $10^{-6}$ eV and $10^{-3}$ eV/Å, respectively. The DFT-D3 method was utilized to address van der Waals interactions [65]. We used WANNIER90 [66] to construct a Wannier-based tight-binding model from the V-d, Sb-p and Sn-p orbitals and further calculated the unfolded bands from the WannierTools package [67]. The lattice constant was used as the experimental value with a=5.5 Å and c=9.3 Å.

## 5. STM measurements

STM measurements were performed with a commercial CreaTec low-temperature system. Single crystals were cleaved *in situ* under cryogenic ultrahigh vacuum at a temperature of approximately 30 K and then immediately inserted into the STM head held at 4.8 K. PtIr tips were calibrated on a single-crystal Au (111) surface prior to the measurements. The spectroscopic data were acquired via the standard lock-in technique at a frequency of 987.5 Hz under a modulation voltage of 2–5 mV.

We typically measure the differential conductance maps dI/dV(r, *eV*) with fields of view of 60 × 60 to 100 × 100 nm$^2$. More than ten local regions have been measured for each doping level. The Lawler–Fujita drift-correction algorithm was used for spectroscopic mapping to remove drift [68], and we used normalized conductance maps $L(r, eV) = dI/dV(r, eV)/(I(r,V)/V)$ in this work to avoid systematic errors related to the set-point effect in constant-current scanning mode [50]. Their FTs, L(q, eV), were rotated to align the lattice direction in the same way for each dataset and then cropped to highlight the main features in the first Brillouin zone. The measurements on strained samples are performed with home-made devices following a previous report [69]. Approximately 0.5% uniaxial tensile strain was applied along one of the lattice directions.

Band folding: In the x=0.09 (0.07) samples, the wavevectors of the QPI signals q' and q'' are approximately 1/3 and 2/3 with respect to the value of $\sqrt{3}/2Q_{Bragg}$ (which can be regarded as the boundary of the q-space "first Brillouin zone") along the Γ–K direction. These long, quasi1D features disperse and are thus related to new scattering between the quasi-1D Kagome orbitals. To explain these features, one needs to fold the original quasi1D band with a wavevector of approximately 1/3 along one of the Γ-M directions, as shown in Fig. 2I. Concurrently, the scattering between the original and folded bands gives rise to a new QPI signal along the related Γ-M direction (green dashed arrows). However, owing to the fussy experimental signature of q' and q'' and the multiple orbital characteristics of the quasi-1D band, we cannot provide an exact value of the scattering vectors solely from the QPI.